# Design of a New Family of Narrow Linewidth Mid-Infrared Lasers

Behsan Behzadi, Mani Hossein-Zadeh, Ravinder K. Jain[*]

*Center For High Technology Materials (CHTM) and Optical Sciences and Engineering (OSE) Program,
Departments of Physics and Electrical Engineering, University of New Mexico, Albuquerque, NM 87131*
*\*Corresponding author: jain@chtm.unm.edu*

**We describe the design of a new family of high spectral brightness narrow linewidth (NLW) mid-infrared (MIR) lasers –- of < 1 MHz anticipated linewidths –- with potential for operation at any target wavelength between 2.5 and 9.5 μm. More specifically, we analyze the potential performance characteristics of mid-infrared distributed feedback (DFB) Raman fiber lasers (RFLs) based on π-phase-shifted (PPS) Fiber Bragg Gratings (FBGs) written in appropriately chosen low-phonon-energy glass fibers. In particular, we calculate anticipated threshold pump powers for optimal laser designs and pump wavelengths for single frequency (fundamental mode) operation of specific mid-infrared DFB-RFLs operating at chosen target wavelengths, and show that these pump powers can be as low as a few milliWatts for MIR DFB-RFLs fabricated with appropriate low-loss small mode area single mode fibers. As such, we clearly establish the PPS-DFB RFL platform as a very practical approach for constructing a broad range of narrow linewidth MIR coherent sources for numerous applications, including proximal and remote sensing of molecules and various high spectral brightness and long coherence length MIR applications.**

## MOTIVATION AND GOALS OF THIS PUBLICATION

Narrow linewidth (NLW) mid-infrared (MIR) sources are needed at specific target wavelengths, particularly in the 2.5 to 9.5 μm range for numerous applications ranging from high spectral brightness and long coherence length applications [1] and nonlinear optics applications [2] to spectroscopic sensing, including remote or trace sensing of targeted molecular species [3-7,11]. Although OPOs (optical parametric oscillators) [7,8] frequently meet the requirements of linewidth, tunability, and power, they are generally cumbersome and expensive, and invariably too large for applications -- such as remote sensing -- that require field-usable portable sources. Semiconductor-based quantum cascade lasers (QCLs) [9-11] have exhibited significant improvements in the last decade, but narrow-linewidth tunable QCLs are primarily based on external cavity configurations [10,12] that are still relatively complicated and cumbersome, and thus relatively expensive as commercial products; moreover, QCLs often require cooling, or only work in the pulsed mode, particularly at shorter mid-IR wavelengths, notably in the 2 to 4.5 μm wavelength range, a spectral region that is highly advantageous for many of the above-cited applications because of: (a) the availability of numerous strongly absorbing molecular transitions in this spectral range [5,6,11] and (b) the ready availability of simpler and compact, low-cost high-sensitivity light detectors in this spectral range (relative to the longer MIR wavelengths). Tunable mid-IR solid state lasers [13,14] based on transition-metal dopants in bulk crystals (Fe:ZnSe and Cr:ZnSe) have also recently been demonstrated in the 3.8 – 5 μm spectral range, but they are relatively bulky and operate at cryogenic temperatures, have limited wavelength coverages, and cumbersome thermal management requirements.

*The goals of this paper are: (1) to theoretically explore the feasibility of demonstrating compact and efficient narrow linewidth mid-infrared sources for the 2.5 to 9.5 μm spectral range based on an alternative technology platform, viz. the Raman fiber laser (RFL) platform; (2) to specify the optimal design parameters of "single mode" or "single frequency" RFLs; and (3) to assess the practicality of constructing RFLs based on such designs in future experimental efforts.*

## RAMAN FIBER LASER BASED NLW-MIR SOURCES: THE PPS-FBG RFL PLATFORM

**Advantages of Raman Fiber Lasers:**
Raman fiber lasers (RFLs) are very attractive sources of laser radiation, particularly at wavelengths at which there are no readily available "invertible" atomic, ionic, or molecular transitions [15-22], as is frequently true at MIR wavelengths. The attractiveness of RFLs stems largely from the fact that stimulated Raman scattering (SRS) with extremely high Raman gains can be easily achieved in low-loss optical fibers at nearly arbitrary wavelengths in the transparency windows of the fibers simply by using moderate pump powers at pump wavelengths corresponding to a relatively broad range of Stokes shifts; this flexibility in the choice of lasing wavelengths and pump wavelengths and pump bandwidths is not only enabled by the broad range of phonon energies available in glasses (because of their amorphous nature) leading to very broad Raman gain bandwidths, but also by the feasibility of using multi-Stokes or "cascading" processes because of the extremely high gains and high conversion efficiencies achievable in each Stokes order [18,19,22]. As such, numerous CW tunable Raman fiber lasers have been demonstrated over the last few decades (since the 1970's) by various researchers

[16-18,20] as highly efficient sources of broadly-tunable coherent radiation at many wavelengths ranging from the visible to the MIR. The high efficiency of Raman fiber lasers –- limited only by the quantum deficit between the pump and the Stokes wavelengths -- is not only enabled by the low fiber losses and the high gains achievable in single-mode optical fibers due to the possibility of confining high pump intensities over long interaction lengths [15-18,21], but also due to the general absence of coupling loss and other superfluous loss sources in "intra-fiber" laser cavities [21,22].

**Broadband Mid-Infrared Fiber Lasers:**
Although efficient mid-IR Raman [19, 20] and rare-earth-doped [23,24] fiber lasers and amplifiers have been demonstrated frequently over relatively broad spectral ranges, the achievement of narrow linewidths in such fiber laser sources is limited in part by the inherent broad gain bandwidths and in part by the extremely high gains and low thresholds of competing processes: namely (1) stimulated Brillouin scattering (SBS) which occurs readily when narrow-linewidth radiation is propagating in very long nearly-single mode fibers [25,26], and by the onset of (2) readily-phase-matched four-wave mixing (FWM) processes [27-30] in long-fiber based laser cavities. As such, the spectral outputs of mid-IR Raman fiber lasers demonstrated in fluoride, tellurite, and chalcogenide fibers [19,20] have been limited to relatively broad linewidths (of the order of a few nm).

**Narrow-linewidth (NLW) Mid-Infrared Fiber Lasers:**
In recent years, narrow-linewidth Raman fiber lasers have been proposed and demonstrated *at near-infrared wavelengths* in silica and germanosilicate fibers [31-35] via the use of advanced distributed feedback (DFB) structures -- notably $\pi$–phase shifted (PPS) Bragg gratings (FBGs) [36] -- which enable strong feedback over a very narrow band of wavelengths (enabling effective discrimination against lasing at wavelengths outside this narrow bandwidth) -- *in relatively short fiber (~ 1 m long) lasers*, circumventing the onset of SBS and FWM processes in such RFLs. In principle, one anticipates that it should be relatively straightforward to extend this PPS-FBG-RFL platform to MIR Raman fiber lasers. *However, this extension to the MIR is not fully obvious due to potential problems related to the availability of appropriate low-loss MIR fibers, the possibility of fabricating appropriate PPS-FBGs, and the need to identify of specific PPS-FBG RFL designs that will allow single-mode NLW operation of such MIR RFLs*; as such, there is no experimental demonstration or theoretical analysis to date on the feasibility and practicality of extending this technology platform to the MIR spectral range. *In this paper, we focus on resolving this need by investigating and quantifying critical parameters that will enable extension of the PPS-DFB RFL source technology platform to MIR wavelengths well beyond 3 microns by using optimized PPS-DFB RFL designs (Section 4) based on advanced mid-IR glass (tellurite, chalcogenide) fibers, including establishing both the requirements on the mode areas and losses in such fibers as well as the optimal requirements for low-pump threshold single frequency PPS-DFB designs. Finally, we describe a roadmap for achieving arbitrary wavelengths between 2.5 and 9.5 microns in the near future, in part by using cascaded MIR fiber Raman lasers as appropriate pumps for such long wavelength MIR NLW sources (Section 5).*

## RELEVANT BACKGROUND DATA ON GLASSES USED FOR MIR FIBERS

**Raman Gain Spectra and Peak Gain Coefficients in Relevant MIR Glasses:**
Table 1 and Fig. 1 summarize the Raman gain properties of key glasses of interest here. Fig. 1 shows the normalized Raman gain spectra, while Table 1 shows (a) the estimated peak Raman gain coefficients [15,37-41] for representative members of each "family" of glasses, (b) the approximate Stokes shifts (in cm$^{-1}$) at the peak Raman gain energies (corresponding to the largest phonon densities), (c) the approximate Raman gain bandwidths (in cm$^{-1}$), and (d) the "nominal" spectral "transparency windows" for several of the commonly used mid-IR glass families (tellurites, chalcogenides, and ZBLAN). The data for silica glass, which has been studied extensively has been added to Table 1 for reference, largely to give perspective on the relative magnitude of the Raman gains, Stokes shifts, and bandwidths in the various mid-IR glasses. Note that most of the Raman gain measurements were performed at near-IR wavelengths (the peak Raman gain in silica at 1.55 micron is about 6.5 ×10$^{-14}$ m/W), and -- except for silica -- the peak Raman gain coefficients in the MIR glasses shown in Table 1 have been scaled to wavelengths near 3 microns by using the inverse wavelength dependence [37] of the Raman gain coefficient on the excitation wavelength, $\lambda_{ex}$ ($G_R \propto 1/\lambda_{ex}$); as such, the peak Raman gain in the relatively "popular" ZBLAN (fluorozirconate) glass -- used extensively for rare-earth-doped MIR FLs [23,24] -- is comparable to (less than a factor of two larger than) that in silica, while the peak Raman gains in several tellurite (e.g., TBZN) and chalcogenide (As$_2$S$_3$, As$_2$Se$_3$) glasses are nearly two to three orders of magnitude larger than those in silica, indicating their strong promise for low-pump-threshold RFLs even with fiber lengths of less than a meter. Note also that the chosen values of Raman gain bandwidths in this table correspond to the 50% gain point in the representative Raman gain curves in Fig. 1; clearly at higher pump powers, much larger gain bandwidths (and a much larger range of corresponding lasing wavelengths of the RFL) should be possible. For instance, a potential Raman gain bandwidth of

**Table 1: Raman gain characteristics of a few key near-IR and mid-IR glasses along with their transparency windows.**

| Glass | Estimated Peak Raman Gain at 3.1 μm (×$10^{-13}$ m/W) | Stokes Shift ($cm^{-1}$) at the Gain Peak | Nominal Raman Gain Bandwidth ($cm^{-1}$) | Transparency Window (μm) | References |
|---|---|---|---|---|---|
| Silica | 0.65 (@ 1.55 μm) | 442 | 200 | 0.35 - 2 | [15,37] |
| Tellurite | 4.5-26 | 741 | 140 | 0.5 - 4.5 | [39,42] |
| Arsenic sulfide ($As_2S_3$) | 21.5-28.5 | 345 | 90 | 1.5 - 6.5 | [40,43] |
| Arsenic selenide ($As_2Se_3$) | 100-255 | 226 | 60 | 1.5 – 9.5 | [38,43] |
| ZBLAN | 0.57-2.1 | 570 | 50 | 0.22 - 4.5 | [41,44] |

750 $cm^{-1}$ (corresponding to Stokes shifts from 75 $cm^{-1}$ to 825 $cm^{-1}$), ie, over 5 times larger than the 140 $cm^{-1}$ value stated in Table 1 (for the 50% gain point), should be readily achievable in the tellurite glass TBZN if the pump powers were high enough to enable attainment of lasing threshold at the 15% gain point in the Raman gain curve (see Fig. 1).

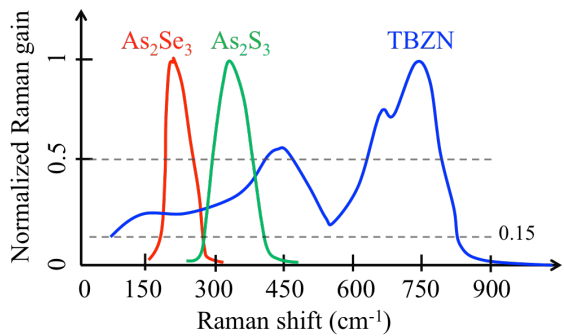

**Fig. 1.** Normalized Raman gain spectra for arsenic selenide ($As_2Se_3$), arsenic sulfide ($As_2S_3$) and tellurite (TBZN) glasses.

**Photosensitivity and Inscription of FBGs in Relevant MIR Glasses**

Another key parameter relevant to the design of DFB-RFLs is the photosensitivity of the core glasses and the maximum value of the refractive index change "modulation", δn, that can be achieved for FBGs inscribed in the cores of the fibers, which in turn determine the coupling coefficient, κ, that may be achievable in appropriate fibers made from such materials. Table 2 depicts information on previously demonstrated values of the index modulation depth (δn) and the corresponding coupling coefficients, along with the fabrication method (illumination wavelengths and writing conditions) chosen for FBGs written on silica (the standard "reference" material) and the above-described mid-IR glasses and silica [45-48].

Table 2 not only shows that that much higher coupling coefficients are readily achievable in FBGs written in **$As_2Se_3$** fibers compared to those written in silica fibers (the technology "baseline" or reference standard), but that for both of the chalcogenide glasses addressed in Table 2, these FBGs can be written quite easily with relatively simple and low power laser sources. On the other hand, if larger Stokes shifts are desired, particularly with the generation of longer MIR wavelengths with relatively short wavelength MIR pump sources, it may be preferable to use the tellurite glass TBZN, in which the FBG lithography process is not as easy as in the chalcogenides; nevertheless this FBG lithography process still compares very favorably to that in silica, both in terms of ease of writing and the magnitudes of the coupling coefficients that are achievable. One should also note that, fabrication of FBGs and the π–phase shift element for longer wavelengths will require lower writing resolutions, simplifying the task of writing such FBGs, particularly for the longer MIR wavelength sources discussed in the cascaded-RFL-pumped NLW sources described in Section 5 below.

Based on the above discussion, it should be clear that the choice of glass to be used for a specific Raman fiber laser application will be based in part on the availability of robust near single-mode fibers -- preferably with very small mode areas -- in these glasses, but also on the magnitudes of the peak Raman gain coefficients, the desired values and ranges of the Raman shifts, and the ease with which the desired PPS-FBGs can be written in these fibers with appropriately large coupling coefficients. In this regard, the selenium-based chalcogenides and the TBZN glasses appear particularly promising, assuming that MIR fibers of sufficiently low mode areas and ultralow losses can be obtained in fibers made from these glasses, the requirements for which are elucidated by our theoretical calculations and related discussion in Sections 4 and 5 below.

Table 2: Previously demonstrated values of refractive index modulation (δn) in various glasses, the corresponding coupling coefficients (κ), and the illumination wavelengths and writing conditions for these demonstrations

| Glass | Achieved values of δn | Estimated coupling coefficient, κ (in cm$^{-1}$) at λ = 3.1 μm | Illumination wavelength (nm) | Writing method and powers/intensity (cw/pulsed: pw, rep rates) used for FBG inscription | Ref. |
|---|---|---|---|---|---|
| Silica | 4 × 10$^{-4}$ | 3.58 | 267 | Near field phase mask, 120fs pulses, 1kHz rep. rate, 8 mW average power and intensity, I = 2.5×10$^{11}$ W/cm$^2$ | [45] |
| Tellurite (TBZN) | 5.4× 10$^{-4}$ | 4.83 | 800 | Near field phase mask, mode-locked Ti:Sapphire laser and w. 1mJ pulse energies @ 1 kHz rep. rate, with a peak intensity, I = 10$^{11}$ (W/cm$^2$) | [46] |
| Chalcogenide (As$_2$S$_3$) | 0.36 × 10$^{-3}$ | 3.22 | 532 | Phase mask technique, 10 mW cw power | [47, 79] |
| Chalcogenide (As$_2$Se$_3$) | 10$^{-2}$ | 89.44 | 633 | Interferometric method, 3 mW cw power | [48, 79] |

## OPTIMIZED DESIGN AND PUMP REQUIREMENTS FOR MIR PPS-FBG RAMAN FIBER LASERS

Fig. 2 schematically depicts the refractive index profile in a phase-shifted FBG structure of total length L written in a fiber aligned for light propagation in the z-direction, with two FBGs (whose index modulation periods Λ have been optimized for operation at the target Stokes wavelength, $\lambda_s$) of lengths L$_1$ and L$_2$ surrounding a central phase shift section of length ΔL (thus L = L$_1$ + L$_2$ + ΔL). As described below, we will focus here primarily on RFLs in which the central phase shift section of length ΔL is assumed to yield a phase shift of π at the target Stokes wavelength, $\lambda_s$, leading to a π–PS or the simpler "PPS" designation.

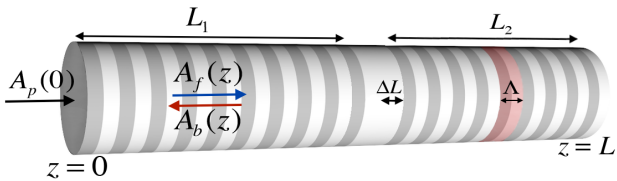

**Fig. 2.** Schematic diagram of fiber core, depicting the terminology for the wave amplitudes used for various sections in our TMM algorithm for the PPS-FBG-RFL model. Λ=λ$_B$/2n$_{eff}$ is the grating period for Bragg wavelength λ$_B$. ΔL is the phase shift section.

In order to determine the optimal PPS-FBG RFL design, and the optimal range of pump powers ignoring longitudinal structural changes, such as changes in the FBG periods and the effective length of the π phase shift section, caused by localized self-phase and cross-phase modulation effects [49,50] for which the mid-IR PPS-FBG RFL may be used as a single mode stable source, we calculate here the threshold pump powers for the fundamental mode $(P_{th,0})$ and the first side mode $(P_{th,1})$ of DFB-RFLs based on chalcogenide (arsenic selenide, As$_2$Se$_3$) and tellurite (TBZN) fibers for various values of the FBG coupling coefficients, κ, FBG lengths, L, fiber losses, α, and effective mode areas, A$_{eff}$.

The theoretical analysis of the PPS-FBG-RFL is most easily done via formulation of the three coupled wave equations for the pump, forward, and backward Stokes waves, as first written for simpler DFB RFLs by Perlin and Winful [51]:

$$\frac{\partial A_P}{\partial z} + \frac{1}{v_P}\frac{\partial A_P}{\partial t} = -\frac{g_P}{2}\left(|A_f|^2 + |A_b|^2\right)A_P$$
$$+i\gamma_P\left(2|A_f|^2 + 2|A_b|^2 + |A_P|^2\right)A_P - \frac{\alpha_{lP}}{2}A_P$$
(1)

$$\frac{\partial A_f}{\partial z} + \frac{1}{v_s}\frac{\partial A_f}{\partial t} = \frac{g_S}{2}|A_P|^2 A_f + i\gamma_s\left(|A_f|^2 + 2|A_b|^2 + 2|A_P|^2\right)A_f$$
$$+i\kappa A_b + i\delta\beta A_f - \frac{\alpha_{ls}}{2}A_f$$
(2)

$$-\frac{\partial A_b}{\partial z} + \frac{1}{v_s}\frac{\partial A_b}{\partial t} = \frac{g_S}{2}|A_P|^2 A_b$$
$$+i\gamma_s\left(2|A_f|^2 + |A_b|^2 + 2|A_P|^2\right)A_b + i\kappa A_f$$
$$+i\delta\beta A_b - \frac{\alpha_{ls}}{2}A_b$$
(3)

where A$_P$, A$_f$ and A$_b$ are the amplitude of pump wave, forward and backward Stokes waves respectively, g$_s$ and - g$_p$ are the stimulated Raman gain and loss coefficients for the laser wavelength λ$_s$ (Stokes wavelength) and the pump wavelength λ$_p$, respectively, and are related to each other by the simple relationship g$_p$ = g$_s$.(λ$_s$/λ$_p$), and $v_P$ and $v_s$ are the phase velocities at the pump and Stokes wavelengths respectively. In these equations, κ is the coupling coefficient between the forward and backward Stokes waves in the FBG, defined by κ = πδn/λ$_B$ for a Bragg wavelength λ$_B$ and a refractive index modulation depth of δn. γ$_s$ and γ$_p$ are the nonlinear propagation constants for the Stokes and pump wavelengths, defined by

$\gamma_s = 2\pi n_2/\lambda_s$ and $\gamma_p = 2\pi n_2/\lambda_p$, where $n_2$ is the Kerr coefficient of the glass. More specially, the terms in the parentheses multiplied by $\gamma_p$ and $\gamma_s$ in Eqs. (1) – (3) are the spatially-averaged contributions of self-phase modulation (SPM) and cross phase modulation (XPM) [51]. Finally, $\delta\beta$ denotes the propagation constant detuning of the laser wavelength (Stokes wavelength) from the Bragg propagation constant and is defined by $\delta\beta = 2\pi n_{eff}(1/\lambda_B - 1/\lambda_s)$, and $\alpha_{ls}$ and $\alpha_{lp}$ denote the linear fiber losses at the laser (Stokes) and pump wavelengths, respectively.

The steady state versions (obtained by setting temporal derivatives to 0) of Eqs. (1) – (3) can be used to determine the threshold conditions at the peak Stokes wavelength for cw operation. As such, assuming a non-depleting pump and the fact that $|A_P| \gg |A_f|, |A_b|$ at the threshold condition, Eqns (2) and (3) simplify to:

$$\frac{\partial A_f}{\partial z} = \left(\frac{g_s}{2}|A_p|^2 - \frac{\alpha_{ls}}{2}\right)A_f + i\left(\delta\beta + 2\gamma_s|A_p|^2\right)A_f + i\kappa A_b \quad (4)$$

$$\frac{\partial A_b}{\partial z} = -\left(\frac{g_s}{2}|A_p|^2 - \frac{\alpha_{ls}}{2}\right)A_b - i\left(\delta\beta + 2\gamma_s|A_p|^2\right)A_b - i\kappa A_f \quad (5)$$

Since the coupled differential Eqs. (4) and (5) are relatively difficult to solve directly, we will apply coupled wave theory and take advantage of the transfer matrix method (TMM)[35,52] for our solutions and threshold pump power computations, as elucidated below. The coupled differential Eqs. (4) and (5) are first converted into matrix form:

$$\frac{d}{dz}\begin{bmatrix}A_f\\A_b\end{bmatrix} = [Q]\begin{bmatrix}A_f\\A_b\end{bmatrix} \quad (6)$$

$$Q = \begin{bmatrix}q & i\kappa\\-i\kappa & -q\end{bmatrix} \quad (7a)$$

$$q = \left(\frac{g_s P_P}{2A_{eff}} - \frac{\alpha_{ls}}{2}\right) + i\left(\delta\beta + 2\gamma_s \frac{P_P}{A_{eff}}\right) \quad (7b)$$

where $P_p$ is the pump power and is related to the pump amplitude, $A_p$, by $P_P/A_{eff} = |A_P|^2$, and $A_{eff}$ is the effective area of the propagating mode inside the fiber. The general solution to Eq. (6) is given by $e^{z[Q]}$, which can be used to find the relation between the forward and backward Stokes waves at $z = L$ and $z = 0$ through the matrix equation:

$$\begin{bmatrix}A_f(L)\\A_b(L)\end{bmatrix} = [T]\begin{bmatrix}A_f(0)\\A_b(0)\end{bmatrix} \quad (8)$$

where:

$$[T] = \left[e^{L2[Q]}\right][P]\left[e^{L1[Q]}\right] \quad (9)$$

and the phase shift matrix $[P]$ is defined by:

$$[P] = \begin{bmatrix}e^{-i\frac{2\pi n_{eff}}{\lambda_s}\Delta L} & 0\\0 & e^{i\frac{2\pi n_{eff}}{\lambda_s}\Delta L}\end{bmatrix} \quad (10)$$

Note that for the case of a perfect π-phase shifted (PPS)-FBG, the P matrix, [P] is given by $\begin{bmatrix}-i & 0\\0 & i\end{bmatrix}$. The analysis for the case of other phase shifts, as may be caused by SPM and PPM are elucidated in related work in progress [50].

The threshold pump powers ($P_{th}$) for the longitudinal modes of the DFB-RFL can then be calculated by applying the boundary conditions of DFB-RFL cavity ($A_b(L) = A_f(0) = 0$), which is equivalent to stating that $T_{22} = 0$. [53]. This threshold pump power required (see matrix Q in Eq. (7) can then be obtained by imposing the requirement of $T_{22} = 0$ and solving for the minimum value of $P_p$ that meets this requirement. We have solved these numerically for symmetric FBGs ($L_1 = L_2$) for a range of values of L via Matlab by using the Newton-Raphson iteration method (for instance, see [54]). Using this approach we start by computing the threshold conditions for oscillation of the fundamental mode of a PSS-DFB-RFL, results for which are plotted below for a broad range of input parameters (κ, L, $g_s$, $A_{eff}$, α). Note that although all the initial calculations below have focused on the case of a specific target wavelengths, namely determining the PPS-FBG-RFL design and pump power requirements for constructing a NLW 3.596 μm laser (designed for sensing of formaldehyde [11] via absorption spectroscopy), the results are relatively independent of these specific wavelength choices, at least to the first order.

Based on the Stokes shifts (see Table 1) in tellurite and arsenic-selenide glasses, the pump wavelengths for optimal pumping of a NLW 3.596 μm laser are chosen to be $\lambda_P = 2.8$ μm and $\lambda_P = 3.3$ μm respectively. For these calculations of the optimal DFB-RFL design and threshold pump values, we have assumed a PPS-FBG of symmetrical design ($L_1 = L_2$) and used core refractive indices of 2.7 and 2.1 and nonlinear refractive indices of $n_2 = 1.1\times10^{-17}$ [55] and $5.0 \times 10^{-19}$ m²/W [56, 39] for arsenic selenide and tellurite fibers respectively. In the mid-IR range (typically near 3.5 μm) commercially available step-index tellurite and chalcogenide fibers have propagation losses as low as 0.25 dB/m. These fibers have relatively large (~100 μm²) mode areas, resulting in relatively large threshold powers for *PPS-DFB RFLs*. However, several researchers have reported photonic crystal (PC) MIR fibers with significantly reduced mode areas, for applications such as efficient supercontinuum generation. More specifically, tellurite and chalcogenide PC fibers with propagation losses of ~ 0.3 dB/m and mode areas < 10 μm² have been reported [57-62]. As such in our calculations we use these numbers as practical values that will be improved in near future. *Note that all of the initial calculations for these DFB RFL designs are based on effective mode areas (A$_{eff}$) of 10 μm$^2$ and propagation loss (α$_{ls}$) of 0.0715 m$^{-1}$ (0.3 dB/m), which correspond to values that are quite easily achievable in these fibers, and are used as our "baseline" fiber design values* [61,62]

Fig. 3 shows the threshold pump power for the fundamental mode ($P_{th,0}$) of the mid-IR DFB-RFL for chalcogenide (arsenic selenide) and tellurite glasses for a range of coupling coefficients that are readily achievable in these glasses (see Table 2). As seen from this figure, for moderate lengths of DFB-RFLs (10 to 12 cm) and moderate coupling coefficients (κ = 0.8 to 1.0 cm$^{-1}$), the estimated threshold pump powers, $P_{th,0}$, are easily achievable (below 80 mW at 3.3 um and 500 mW at 2.8 μm for the As$_2$Se$_3$ and TBZN RFLs respectively) with readily available broadband MIR pump lasers. Also, as expected, the threshold pump power of the fundamental mode decreases significantly by increasing the FBG length ($L$) and coupling coefficient (κ) approximately at a rate of $\kappa e^{-\kappa L}$ [63].

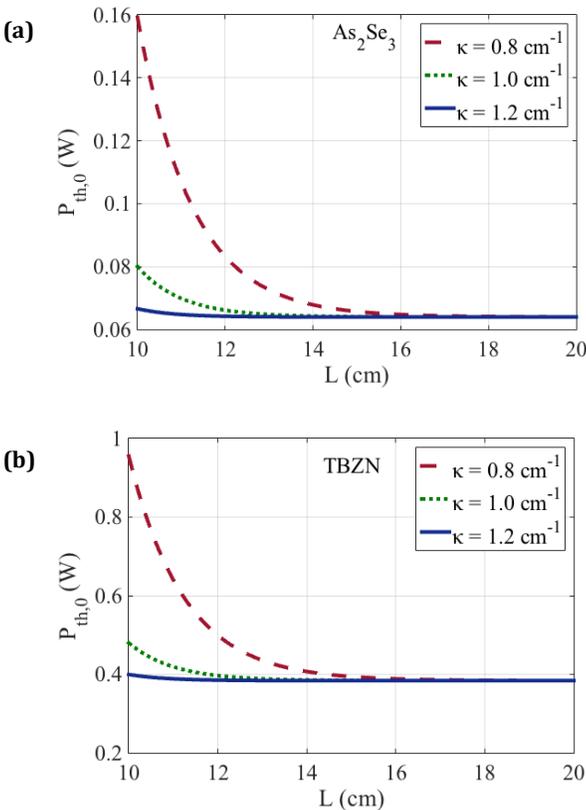

**Fig. 3.** Threshold pump powers for the fundamental mode ($P_{th,0}$) plotted as a function of the FBG length for (a) chalcogenide (As$_2$Se$_3$) and (b) tellurite (TBZN) fibers, assuming a baseline fiber design with an effective mode area (A$_{eff}$) of 10 μm$^2$ and a propagation loss (α) of 0.3 dB/m. Three "nominal" but highly realistic values of coupling coefficient (κ) have been chosen for three independent plots for each case.

Note that for a given value of κ, $P_{th,0}$ reaches a minimum value ($P_{th,0-m}$) at certain FBG length ($L_{max}$) above which fiber loss dominates the total loss (internal + external) in the cavity. This threshold value is approximately given by $P_{th,0-min} \approx \alpha_{ls} \times A_{eff}/g_s$ [64]. As an example, for an arsenic-selenide DFB-RFL with a coupling coefficient κ = 1 cm$^{-1}$, the threshold pump power does not become smaller than $P_{th,0-m}$ = 64 mW for lengths above $L_{max}$ = 18 cm, indicating a very important and specific design parameter. Although one may use $L \geq L_{max}$ as the preferred DFB-RFL length, say for increased output power and increased stability, the need for single mode operation (and avoidance of frequency noise and self-pulsing [65]) imposes a further limit on $L$.

Single mode operation of such DFB-RFLs requires avoidance of lasing at higher order modes, and limits the maximum pump powers that may be used for optimized NLW operation. This issue is clarified by considering the onset of lasing at the "first side mode" frequency of the PPS-FBG, whose spectral shift (relative to the lowest order mode) and spatial intensity distribution are illustrated in Fig. 4 below. Fig. 4(a) shows the transmission spectrum for a near symmetric PPS-FBG; here $m$ = 0 and $m$ = 1 correspond to the lowest order longitudinal modes, referred to as the fundamental mode (narrowest linewidth) and the first side mode respectively. Fig. 4(b) schematically illustrates the optical intensity distribution along the length of the cavity (PPS-FBG) for the fundamental mode ($m$ = 0) and the first side mode ($m$ = 1); as seen in Fig. 4 (b), the intensity of the fundamental mode is highly concentrated in the center of the phase shift section, while the maximum intensity of the first side mode is located near the center of the two FBGs surrounding the phase-shift section [65].

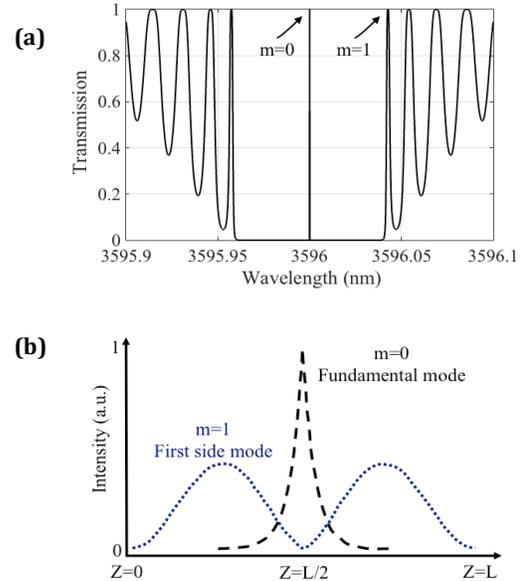

**Fig. 4.** (a) Transmission spectrum of a symmetric 15 cm long π-phase-shifted FBG for λ$_b$ = 3.596 μm, (with κ= 0.5 cm$^{-1}$). (b) Schematic distribution of the optical intensity as a function of distance for the fundamental (m=0) and the first side mode (m=1) of for such a symmetric DFB-RFL (with a π-phase shift section centered at z = L/2).

In order to establish the upper limits of the pump powers that may be used while maintaining single-mode (ie, single wavelength) operation, we have calculated and plotted the threshold pump powers for the first side mode ($P_{th,1}$) by using an initial value of the phase offset, $\delta\beta$, from Fig 4 (a) and finding the higher order solutions for T$_{22}$ (P$_p$, $\delta\beta$) = 0. As seen in Fig. 5, the behavior of $P_{th,1}$ (as a function of $L$) is qualitatively similar to that of the $P_{th,0}$, albeit at significantly higher values of pump powers, the consequences of which are elaborated in the next paragraph by considering the role of the "power budget", $\Delta P_{th} = P_{th,1} - P_{th,0}$ on the design of single mode RFLs.

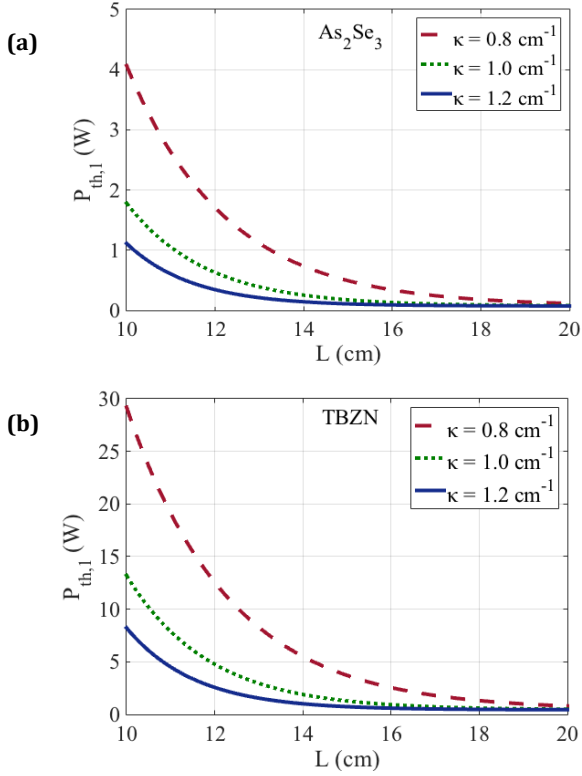

**Fig. 5.** Threshold pump powers ($P_{th,1}$) for the first side modes as a function of the FBG length for the "baseline fiber designs", ie the same choices of mode areas ($A_{eff}$ = 10 µm²), fiber loss ($\alpha$ = 0.3 dB/m), and for the same coupling coefficients ($\kappa$) as those used for the fundamental mode calculations (see Fig. 3), ie for identical (a) chalcogenide ($As_2Se_3$) and (b) tellurite (TBZN) fiber-based DFB-RFLs.

Clearly for single-mode operation, the pump power ($P_p$) should be larger than $P_{th,o}$ and smaller than $P_{th,1}$. In addition, $\Delta P_{th}$ (=$P_{th,1}$ - $P_{th,0}$) has to be large enough to facilitate emission of significant output power during single mode operation. The required conditions for optimal DFB-RFL designs and pumping conditions are elucidated in Fig. 6, which illustrates $\Delta P_{th}$ in the form of color-coded contour plots in κ-L space for the (a) arsenic selenide (a) and tellurite (b) DFB-RFLs ($A_{eff}$ =10µm² and $\alpha$=0.3 dB/m) of Figs. 3 and 5. The dashed lines correspond to values of $P_{th,0}$ = 100 mW (for AsSe) and 1 W (for TBZN), while the solid lines correspond to absorption-limited threshold powers for these two fiber lasers ($P_{th,0-min}$).

Based on Fig. 6, for a specific $P_{th,0}$, a shorter FBG length and a larger coupling coefficient results clearly in a larger value of $\Delta P_{th}$ (i.e. a larger single mode operation range). As an example, for an arsenic selenide DFB-RFL with a $P_{th}$ of 100 mW (Fig. 6 (a), dashed line), changing the DFB-RFL design parameters from $L$ = 7 cm and κ = 1.1 cm⁻¹ to $L$ = 25 cm, κ = 0.3 cm⁻¹, reduces the power budget $\Delta P_{th}$ from 8 Watts to 3 Watts, making the former design option (shorter DFB-RFL with a larger coupling coefficient) much more desirable for better single mode performance and higher output powers with minimal modal noise. It is also obvious that pump power budgets of over 50 Watts -- and *relatively high (Watt level) output powers with ultra-narrow output linewidths* -- should be readily achievable with DFB-RFLs of a few cm length and nominal coupling coefficients while keeping the threshold pump powers to less than 1 Watt. Finally, for chalcogenide DFB-RFLs much higher laser output powers –- and larger power budgets –- should also be possible with the use of single mode fibers with much larger mode areas with the use of higher pump powers, limited primarily by the damage threshold of the fiber.

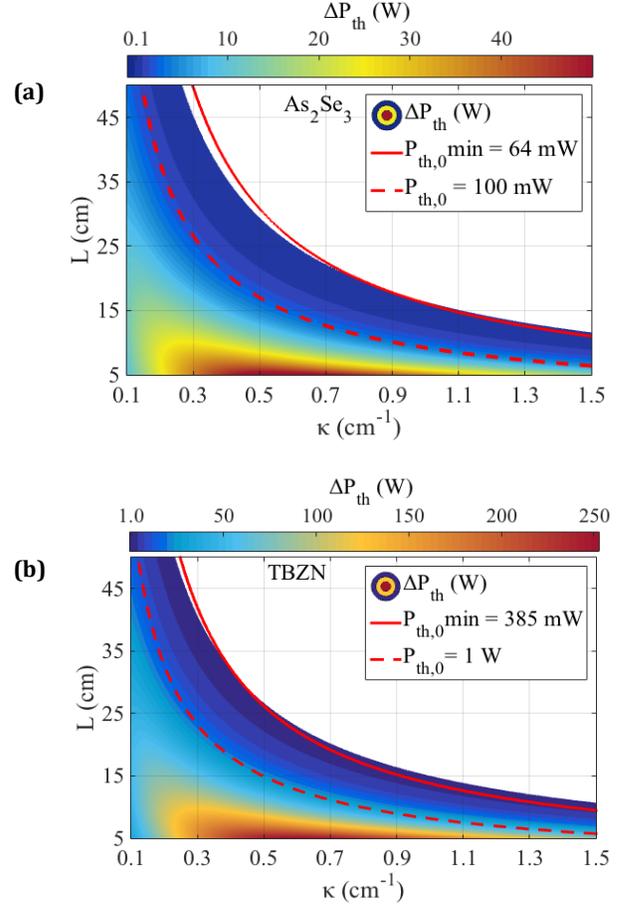

**Fig. 6.** $\Delta P_{th}$ contours (color coded) plotted in κ - L space for the (a) arsenic selenide and (b) tellurite DFB-RFLs of Figs. 3 and 5. The dashed lines correspond to $P_{th,0}$ values of 100 mW (AsSe) and 1 W (TBZN), while the solid line corresponds to the absorption limited threshold powers ($P_{th,0-min}$) of 64 mW and 385 mW respectively. As for the RFLs in Figs. 3 and 5, $A_{eff}$= 10 µm² and $\alpha$=0.3 dB/m.

The above-described behavior can also be explained qualitatively by considering the power distribution of the fundamental and first side modes in DFB-RFLs (Fig.4b)[65]. For the fundamental mode, the power is distributed symmetrically with its maximum around the middle of PPS-FBG (ie the position of π-phase shift position) while for the first side mode, the power is distributed symmetrically with a minimum at the middle and the two maxima offset significantly from the center of the PPS-FBG. Thus decreasing the FBG length (while increasing the coupling coefficient) causes the nonlinear gain for the first side mode to diminish much more rapidly than that of the fundamental mode, enabling higher modal discrimination and better single mode performance. As a complementary point, increasing the FBG length for a fixed coupling coefficient, say from 5 cm to 50 cm, will increase the gain and thus decrease the threshold pump power of the first side mode much faster than that of the fundamental mode. However, for long DFB-RFL lengths, the gain is offset by the increased net loss, and further increases in the fiber length will not change the pump thresholds and power budgets, $\Delta P_{th}$, significantly. As such, operation of the DFB-RFL at its absorption limited fundamental mode threshold ($P_{th,0-m}$) will limit its single-frequency operation, but operating it at pump powers that are about a factor of two above its $P_{th,0-m}$, will increase the side mode threshold and side-mode suppression significantly.

To ascertain the optimum values of $L$ and $\kappa$ from a different perspective, we consider the "modal discrimination ratio" $M = \Delta P_{th}/P_{th,0}$ as a figure of merit, and make contour plots of this dimensionless ratio in $\kappa$-$L$ space for our "baseline" arsenic selenide DFB-RFLs, as shown in Fig. 7. This figure clearly elucidates a strong preference for using large coupling coefficients (and short DFB-RFLs) for robust single mode operation, and that values of M > 30 should be readily achievable for this "modal discrimination figure of merit".

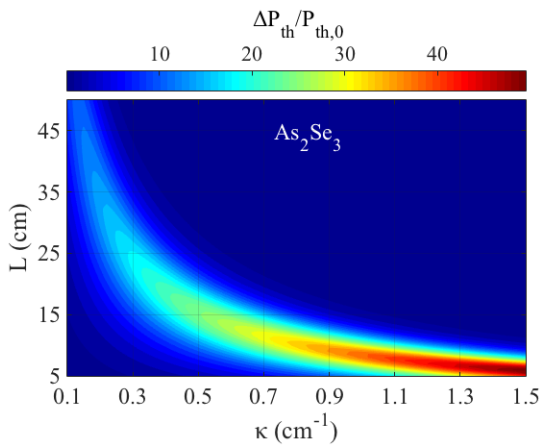

**Fig. 7.** Contours of the modal discrimination ratio, $M = \Delta P_{th}/P_{th,0}$ in $\kappa$-$L$ space for our baseline arsenic selenide fibers (with $\alpha$=0.3 dB/m and $A_{eff}$= 10 µm²).

The threshold pump powers of single-mode DFB-RFLs can be reduced significantly not only by using large coupling coefficients, but also by using fibers with much lower intrinsic losses and smaller mode areas. Fabrication of low-loss, ultra-small effective mode area mid-IR fibers (in particular, those based on tellurite and arsenic selenide glasses) is a challenging task. However this challenge has been partly addressed by recent developments in highly nonlinear mid-IR arsenic sulfide and tellurite photonic crystal fibers, such as those used for supercontinuum generation [57-62]. Arsenic selenide fibers with effective mode areas of below 3.5µm² at $\lambda$= 3 µm have been demonstrated [57]. With the use of higher purity glasses and improved fabrication techniques, fiber losses of less than 0.1 dB/m over broad spectral ranges in the MIR are anticipated in the foreseeable future. As such, we have evaluated the impact of lower losses and a range of effective mode areas on the performance of mid-IR DFB-RFLs. In particular, we have plotted the fundamental mode thresholds for $\kappa$= 1.0 cm⁻¹ and L=10 cm as a function of fiber loss for four different effective mode areas in Fig. 8.

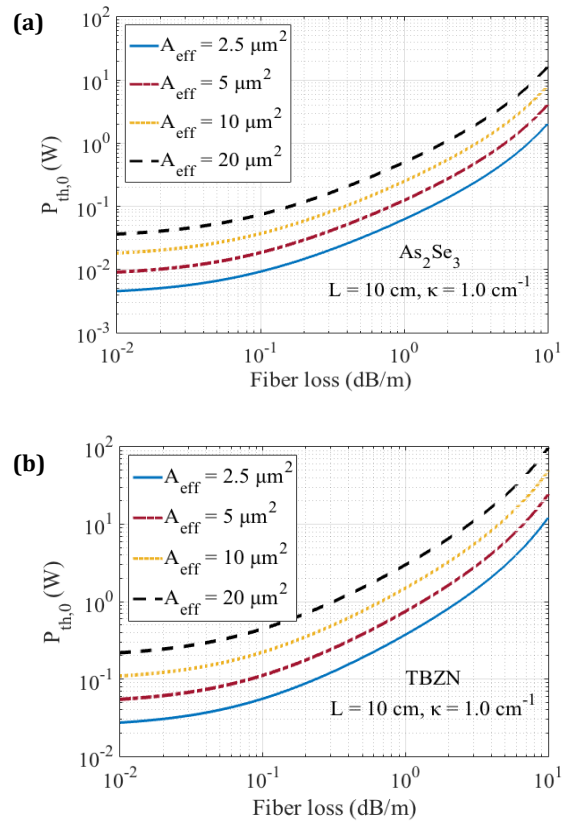

**Fig. 8.** Fundamental mode threshold powers ($P_{th,0}$) plotted as a function of fiber loss (from 0.01 to 10dB/m) for (a) arsenic selenide (As₂Se₃) and (b) tellurite(TBZN) fibers of four different mode areas ($A_{eff}$): 2.5 µm², 5 µm², 10 µm² and 20 µm². For these calculations, we have assumed that $\lambda$= 3.6 µm, $\kappa$=1.0 cm⁻¹, and L = 10 cm.

As depicted in Fig. 8, threshold pump powers below 10 mW and 100mW can be achieved with small mode-area DFB-RFLs made of ultra-low loss chalcogenide and tellurite fibers respectively. Although presence of defects and impurities introduced during fabrication process results in relatively large propagation loss in currently available mid-IR fibers (compared to those in NIR), the intrinsic material losses are much smaller and much lower loss fibers (<0.05 dB/m) are anticipated in the foreseeable future.

We have also studied the impact of the operating wavelength on the threshold pump power. Fig. 9 shows the anticipated wavelength dependence of the estimated fundamental mode threshold of a As₂Se₃ DFB-RFL for $\kappa$=1.0 cm⁻¹, L=10 cm and two effective areas. The increase in the pump threshold is caused by the projected decrease in the Raman gain and the mode intensity -- as well as increased fiber losses -- at longer wavelengths (i.e. about 0.3 dB/m at 2.5 µm and 0.7 dB/m at 9.5 µm); nevertheless our calculations show that narrow-linewidth DFB-RFLs at wavelengths as long as 10 microns should be achievable with the use of appropriate pump sources.

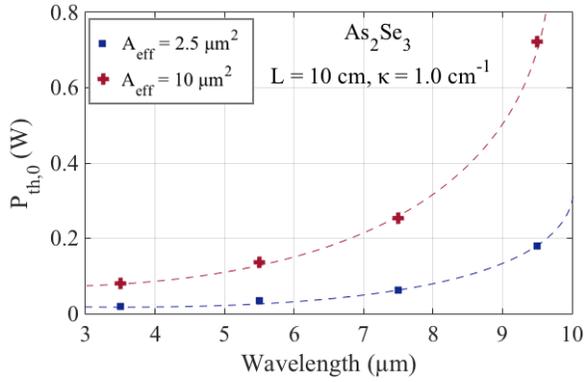

**Fig. 9.** Estimated fundamental mode threshold of a As$_2$Se$_3$ DFB-RFL plotted as a function of the operating wavelength for $\kappa$=1.0cm$^{-1}$, L=10 cm and A$_{eff}$=2.5 µm$^2$ and A$_{eff}$=10 µm$^2$. Fiber losses are taken from [61], [62].

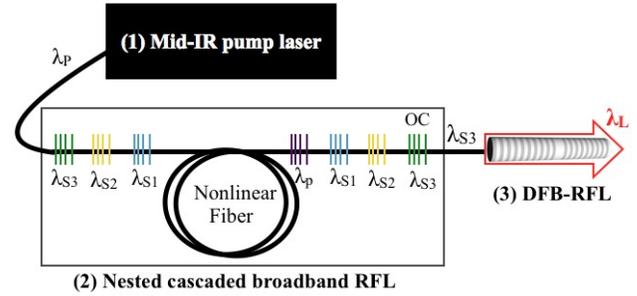

**Fig. 10.** A proposed schematic arrangement for a mid-IR Raman fiber laser system based on three (n = 3) nested cascaded broadband RFLs to generate pump broadband pump wavelengths as long as 5 µm, and narrow linewidth laser radiation at any arbitrary wavelength up to 9.5 µm.

## ANTICIPATED SPECTRAL COVERAGE OF NLW MIR SOURCES BASED ON PPS-DFB-RFLs

Our discussion in Section 4 above focused on the optimization of the design of NLW MIR sources based on PPS-FBG-RFLs, with particular attention given to RFLs based on specific tellurite (TBZN) and chalcogenide (As$_2$Se$_3$) glass systems, with a primary focus on optimal DFB-RFL designs and pump requirements for a NLW 3.6 micron laser (estimated linewidth < 1 MHz) based on these assumptions. It is very clear that the above-stated principles apply equally well to a much broader range of mid-IR glass materials, including fluorides and other tellurite and chalcogenide glasses, and to other MIR glasses and fibers whose Raman gains, Stokes shifts, FBG imprinting capabilities and range of achievable DFB coupling coefficients, loss coefficients at the pump and the target NLW wavelengths and mode areas are clearly specified. As such, it is clear that these arguments can be extended to the generation of high-quality NLW MIR sources at near-arbitrary MIR wavelengths, provided that fibers with sufficiently low losses and appropriate pump sources are achievable.

There are clearly a broad diversity of high power broadband MIR pump sources cited in the literature [66-68], and most of these can be used with appropriate beam shaping and fiber coupling lens assemblies as pumps for the proposed RFLs, particularly since the pump bandwidth requirements are relatively flexible (bandwidths as large as a few nm are acceptable without significant loss in conversion efficiency due to the large Raman gain bandwidths of the glasses used). Nevertheless, for simplicity we focus here first on the possible spectral coverage that may be achievable with the use of demonstrated and easily-buildable fiber lasers of pulse durations longer than about 10 ns (depending on the DFB length, to justify the steady-state analysis requirement and enable sufficient reduction in the linewidth), including 2 µm Tm:SiO$_2$ FLs [69], 3 µm Er: ZBLAN FLs [23, 24, 70] and conventional "broadband" cascaded fiber Raman lasers [21, 71]. Some specific examples of potential sources are illustrated schematically in Figs. 10 and 11 below, with key issues elucidated in the discussion in the following paragraphs.

Fig. 10 schematically depicts a particularly versatile "nested RFL" pump arrangement that may be used to generate a broad range of fiber-laser-based pump sources to generate high power pump lasers at various MIR pump wavelengths (up to 5 µm), thereby enabling PPS-FBG-RFL based NLW MIR sources at any target wavelength as long as 6.5 µm. The system consists of three stages: (1) a stable and high power mid-IR pump laser; (2) a pump "wavelength shifter" (consisting of a nested cavity cascaded multi-Stokes Raman fiber laser) whose high power broadband emission output is one Stokes shift away from the target wavelength; and (3) the NLW mid-IR PPS-FBG-RFL designed for the specific target NLW MIR emission wavelength. In Fig. 11, the choice of high power pump fiber lasers such as Er:ZBLAN (upto 24 W pump power, $\lambda_p$ = 2.7 - 3.0 µm)[23, 24, 70] and/or higher power Tm:silica fiber lasers (upto 250W pump power, $\lambda_p$ = 1.9 - 2.1 µm) [69] is based on currently demonstrated high power stable MIR fiber laser sources. In this "nested cascaded RFL" scheme, each of the "intermediate" Stokes wavelengths $\lambda_{si}$ (in the example in Fig. 10, i = 1, 2) oscillates in the nested cavity with broadband high reflectivity FBGs without significant outcoupling, while the final Raman cascaded wavelength $\lambda_{s,n}$ ($\lambda_{s,3}$ in Fig. 10, ie, n = 3) outcouples with an appropriately high coupling efficiency to generate a high power pump source at $\lambda_{s,n}$, with a power conversion efficiency limited primarily by the quantum deficit between the original MIR FL pump and the final Stokes wavelength. With the use of higher power pumps (for instance, see [68]), large Stokes shift broadband tellurite RFL-based pumps, and a larger number (n > 6) of "nested" cascaded RFL pumps [18, 22], this nested RFL-based concept is easily extendible to NLW sources upto 10 µm or so, limited only by the availability of small-mode-area chalcogenide fibers with sufficiently high nonlinearities and transparencies. This pump scheme is particularly important if one needs to extend the narrow linewidth RFL source to specific target long MIR wavelengths such as those corresponding to specific absorption lines in target molecules for trace or remote detection systems. Note that the threshold pump powers and conversion efficiencies for each Stokes order can be decreased by simply increasing the nonlinear fiber (Raman gain medium) length, as long as the fibers exhibit sufficiently low losses.

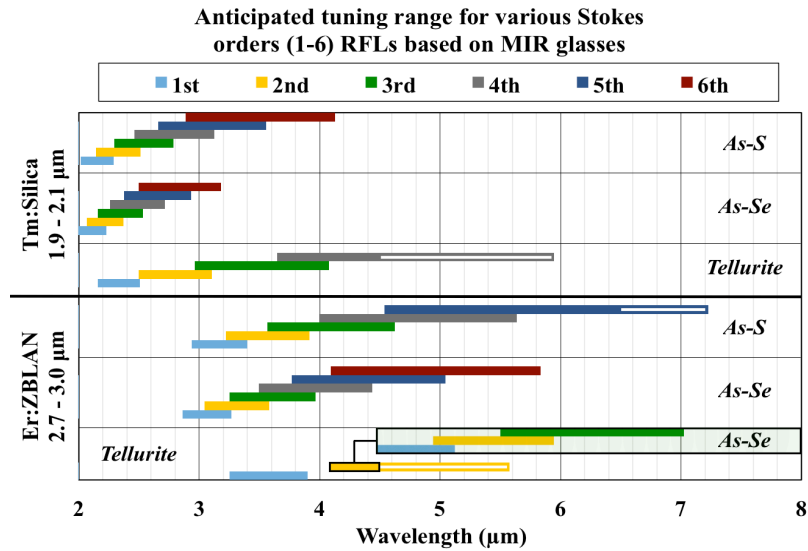

**Fig. 11:** Ranges of cascaded Raman wavelengths for various Stokes orders in specific mid-IR glass-based RFLs pumped by Er:ZBLAN and Tm:silica fiber lasers. The solid rectangles (or bars) indicate wavelength ranges of efficient Raman conversion, and the hollow rectangles correspond to wavelength ranges that may be limited by transparency limits of specifically chosen glasses (approx. 4.5 µm for most tellurites and 6.5 µm for $As_2S_3$). The last (6th) row corresponds to the use of both tellurite and $As_2Se_3$ based RFLs, with 2 orders of nested tellurite fiber-based RFLs followed by 3 orders of nested cascaded $As_2Se_3$ RFLs (inside the large green rectangle) to yield pump wavelengths as long as 7 µm to pump NLW $As_2Se_3$ DFB-RFLs at target NLW wavelengths between 7 and 9.5 µm.

As such, the nonlinear fiber, the power and wavelength of the pump source and the Bragg wavelengths and reflectivities of the FBGs are designed for efficient generation of broadband power at a specific wavelength, and its subsequent delivery to the final narrowband DFB-RFL stage. Fig. 11 shows a few representative examples of two primary pump lasers (Tm:silica and Er:ZBLAN fiber lasers) for pumping the broadband nested RFL cascade, and conservative estimates of the anticipated wavelength coverages that should be possible with various orders in the RFL cascade based on the Raman gain spectrum of arsenic selenide (210 - 270 $cm^{-1}$), arsenic sulfide (300 - 390 $cm^{-1}$) and tellurite (630 - 770 $cm^{-1}$) glasses. As a specific example, one can choose a > 20 Watt Er:ZBLAN pump laser wavelength at 2.78 micron and design the arsenic selenide fiber-based nested cavity RFL by choosing HR FBGs at $\lambda_{S1}$ = 2.98 µm, $\lambda_{S2}$ = 3.32 µm to match the Raman gain peak in arsenic selenide, and generate efficiently out-coupled 3.32 µm laser radiation in the 3rd Stokes order to pump a PPS-FBG-RFL designed for narrow linewidth emission at 3.596 µm. Note that such a multi-Stokes Raman fiber laser could be easily operated with a net conversion efficiency [18, 22, 71] of over 40%, corresponding to NLW emission at Watt-level output powers and extremely high spatial and spectral brightness characteristics. Likewise, multi-Watt output power pumps are anticipated at 4.5 µm in the system depicted in the 6th row of Fig. 11 to enable power levels as high as 100 mWatts at wavelengths as long as 10 µm.

## LINEWIDTH and WAVELENGTH TUNABILITY

**Linewidth Issues:**

Linewidths as small as 10 kHz have been demonstrated in NIR silica fiber based DFB-RFLs (of 30 cm length) [72, 34], and linewidths < 10 MHz have been demonstrated in short (12.4 cm long) NIR DFB-RFLs based on relatively lossy germanosilicate fibers [33]. Note that for ultralow loss PPS-DFBs (such as at NIR wavelengths near 1.55 µm ) the linewidth for $\kappa L \gg 1$ is estimated to be given approximately by $\sim 4\kappa e^{-\kappa L}/2\pi n_{eff}$ [73]. However, in order to estimate upper limits for the linewidths of the proposed MIR DFB RFLs, we have calculated the cold cavity linewidths using the transfer matrix [T] for $P_p = 0$ (see Eq. 9)... Fig. 12 shows the calculated values of the anticipated cold cavity linewidths in a PPS-DFB made of AsSe as a function of fiber loss for several values of $\kappa$ and L. In the low loss limit [73], the estimated linewidths for MIR DFB-RFLs have an inverse exponential dependence on $\kappa L$. Emission linewidths of less than 1 MHz are anticipated when the fiber loss is less than 0.3 dB/cm (as shown in Fig. 12 for $\kappa$ = 1 $cm^{-1}$ and L = 10 cm). Even a linewidth of a few MHz (anticipated for a broader range of DFB-RFLs design parameters) is sufficient for most MIR spectroscopic and sensing applications.

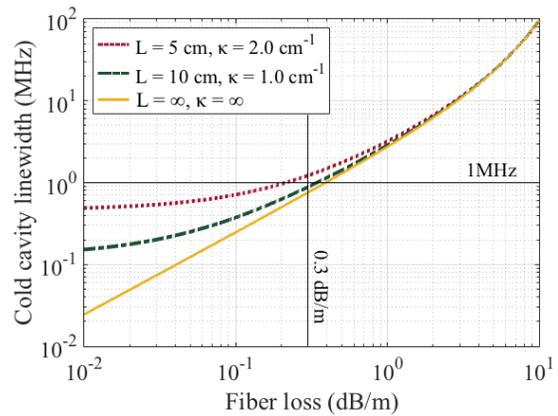

**Fig.12.** Estimated linewidths for $As_2Se_3$ DFB-RFLs as a function of the fiber loss.

The limit for infinitely long PPS-DFB and infinitely large coupling coefficient indicates that fiber-loss limited linewidth for available mid-IR fibers is larger than 100 kHz (note that for silica based PPS-DFBs at 1.55 µm, the loss-limited linewidth is estimated to be of the order of a few kHz).

Note that when nonlinear optically or thermally induced phase shifts are negligible (see Section 6 below), the actual linewidth of PPS-FBG-RFL can be smaller than that of the "cold cavity", as elaborated in [50].

**Wavelength Tunability:**
The emission wavelength of the DFB-RFLs can be tuned by stretching the entire structure, say via a mechanical beam bending method [74] or via various linear FBG "stretching or squeezing" assemblies such as stretching of an appropriately long PZT structure [75,76] to which the FBG is firmly attached. As such, continuous tunabilities of the order of 1 nm (or several 10's of GHz), sufficient for most spectroscopic sensing applications, should be easily achievable.

## POWER SCALING ISSUES

In the discussion above, we have projected power levels of the order of Watts in tellurite based mid-IR DFB-RFLs. We now address the issue of anticipated power scaling limits in future MIR DFB-RFLs. We anticipate the use of "downstream" high gain amplifiers, such as broadband electrically-pumped inverted molecular gas amplifiers or Raman fiber amplifiers (RFAs) to facilitate boosting power levels from the milliWatt levels to Watt levels (the latter -- RFAs -- are limited by the constraints imposed by the onset of SBS in such Raman amplifiers) as the best method of achieving high power NLW radiation from such sources, as opposed to achieving Watt level powers directly from the NLW DFB-RFLs, because of constraints imposed by nonlinear optical and thermal effects and materials' damage thresholds, as elucidated below.

**Nonlinear Optical and Thermal Effects:**
Nonlinear optical and thermal Effects can impose relatively severe constraints on the operation of PPS-FBG-RFLs at high powers [50]. As an example, at high pump or lasing powers, the circulating MIR power in the central π-phase shift region or even in the FBG can be large enough (see Fig. 4 (b)) to cause significant refractive index changes due to the Kerr effect or due to thermo-optic effects (if the absorption coefficients are large enough to cause significant increases in the local temperature of the core glass). Such effects can not only cause a significant deviation in the magnitude of the phase shift in the central region from its optimal value of π at the target wavelength, but also cause a change the effective grating period, resulting in a strong reduction of the quality factor in the resonance for the fundamental mode. Such effects can severely limit the maximum stable single mode output power that can be extracted from PPS-DFB-RFLs [49,50]).

**Materials Damage Thresholds:**
Several of the low-phonon energy mid-IR glasses that exhibit large Raman gains -- notably the chalcogenides -- suffer from relatively low values of optical intensities for their surface damage thresholds, in part because of their large expansion coefficients and low glass transition temperatures ($T_g$). Chalcogenide glasses have been reported to exhibit surface damage at intensities of 3 GW/cm$^2$ in pulsed laser studies [77] and to exhibit damage thresholds as low as 5 MW/cm$^2$ for cw laser radiation. The surface damage thresholds for tellurite glasses are estimated to be of the order of 35 MW/cm$^2$ for cw irradiation; for some tellurite glasses such as TBZN and TPZN, the damage thresholds are significantly higher, and measured to be as high as 20 GW/cm$^2$ (under pulsed laser irradiation) [42], and much higher damage thresholds are expected for tungsten-tellurite glasses such as TWL [42,78]. Note that the optical damage threshold intensities depend strongly on excitation wavelength and beam quality; either way, these damage threshold values are anticipated to increase significantly with further research and development of optimized MIR glasses and optical fibers based on significantly improved material quality, notably those based on glasses with fewer impurity ions, lower defect densities, reduced attenuation coefficients, and improved heat treatment and surface preparation, all of which are anticipated to be natural trends in the materials and fiber developments of such MIR fibers in the foreseeable future. Nevertheless, the measured optical damage threshold intensities establish a preliminary estimate, or a "guideline value" of an upper limit for the maximum pump powers that may be used, and elucidate the need to promptly and carefully measure these values at the desired pump excitation conditions (wavelength, beam quality, and pulse duration) for specific fibers that are developed or acquired for use in such RFLs and RFAs, prior to the initiation of the fabrication and testing of these devices. It should also be noted that the reported damage thresholds for pulsed laser irradiation indicate that optical damage issues may not be a significant limitation for generation of multi-Watt peak power low duty cycle transform-limited narrow linewidth pulses of durations of the order of 10 to 100 nanoseconds.

## SUMMARY

In summary, we have described the design of a new family of high spectral brightness narrow linewidth (NLW) mid-infrared (MIR) lasers –- of < 1 MHz anticipated linewidths –- with potential for operation at any target wavelength between 2.5 µm and 9.5 µm. More specifically, we have analyzed the design and potential performance characteristics of mid-infrared distributed feedback (DFB) Raman fiber lasers (RFLs) based on π-phase-shifted (PPS) Fiber Bragg Gratings (FBGs) written in mid-IR fibers based on appropriately chosen low-phonon-energy glasses, such as chalcogenides and tellurites. We have shown that such lasers can be realized in a practical manner in the near future because of recent developments and promising advances in mid-IR fiber and fiber Bragg grating inscription technology, and anticipated that significant improvements in the performance of such devices should occur in the near future, in part due to anticipated developments and advances in the core mid-IR fiber technologies.

We have also calculated and estimated threshold pump powers for specific PPS-FBG-RFL laser designs and pump wavelengths for single frequency (fundamental mode) operation at chosen target wavelengths, focusing initially on the specific example of spectroscopic sensing of formaldehyde via its well-known absorption feature at 3. 6 µm. We show that the threshold pump powers for these -- and other similar – DFB-RFLs can be as low as a few milliWatts ( < 50 mW) for optimized devices fabricated with appropriate low-loss small mode area single mode fibers. We have also analyzed and established the feasibility of achieving arbitrary wavelengths between 2.5 and 9.5 µm in the near future, in part by using cascaded MIR fiber Raman lasers as appropriate pumps for such long wavelength MIR NLW sources. As such, we clearly establish that the PPS-

DFB RFL platform is a very practical approach for demonstrating new NLW MIR lasers in the near future, and that this platform should result in an exciting new family of narrow linewidth MIR coherent sources for numerous applications, including proximal and remote sensing of molecules at trace levels and other high spectral brightness and long coherence length MIR applications.

**Funding Information**. National Science Foundation, NSF Grant # 1232263; Army Research Office, ARO Grant # W911NF-13-1-011.

**Acknowledgment**. We thank M. Aliannezahedi and M. Klopfer for assistance in research and performing preliminary computations for many of the results reported here.